\def\'#1{\ifx#1i{\accent"13\i}\else{\accent"13#1}\fi}
\def\alamenos#1{$^{-#1}$}

\def\diezala#1{10$^{#1}$}

\newcommand\msun{\rm M_{\odot}}

\newcommand\be{\begin{equation}}
\newcommand\en{\end{equation}}

\documentclass[preprint]{rmaa}
\usepackage{paralist}
\usepackage{psfrag,color}

\SetYear{2006}
\SetConfTitle{-}

\title{Remarks on Rapid vs. Slow Star Formation} 

\author{
  Javier Ballesteros-Paredes\altaffilmark{1} 
  Lee Hartmann\altaffilmark{2}
}

\altaffiltext{1}{Centro de Radioastronom\'\i{}a y Astrof\'{\i}sica,
UNAM, Morelia, M\'exico.} 
\altaffiltext{2}{University of Michigan, Michigan, USA.}

\shortauthor{Ballesteros-Paredes \& Hartmann}
\shorttitle{Rapid vs. slow star formation}

\fulladdresses{
\item Javier Ballesteros-Paredes: Centro de Radioastronom\'\i a y
Astrof\'\i sica,  Universidad Nacional Aut\'onoma de M\'exico, Campus
Morelia,  Apartado Postal 3--72, 58090 Morelia, Michoac\'an, M\'exico
(\email{j.ballesteros@astrosmo.unam.mx}).  
\item Lee Hartmann: Department of Astronomy, University of Michigan,
825 Dennison  Building, 500 Church Street, Ann Arbor, MI 48109
(\email{lhartm@umich.edu}) 
}

 \abstract{We discuss problems with some observational estimates
 indicating long protostellar core lifetimes and large stellar age
 spreads in molecular clouds.  We also point out some additional
 observational constraints which suggest that protostellar cores do
 not have long lifetimes before collapsing.  For external galaxies, we
 argue that the widths of spiral arms does not imply a long
 star-formation process, since the formation of massive stars will
 disrupt molecular clouds, move material around, compress it in other
 regions which produce new star-forming clouds.  Thus, it seems
 unavoidable that this cyclical process will result in an extended
 period of enhanced star formation, which does not represent the
 survival time of any individual molecular cloud.  We argue that the
 rapid star formation indicated observationally is also easier to
 understand theoretically than the traditional scenario of slow
 quasi-static contraction with ambipolar diffusion.}

 \resumen{Discutimos problemas con algunas estimaciones
 observacionales que indican tanto tiempos de vida largos para los
 n\'ucleos densos protoestelares, como grandes dispersiones de las
 edades de las estrellas asociadas a nubes moleculares.  Notamos
 algunas restricciones observacionales que sugieren que los n\'ucleos
 densos no tienen tiempos de vida largos antes de colapsar.  Para
 galaxias externas, argumentamos que los anchos de los brazos
 espirales no implican un proceso de formaci\'on estelar largo, ya que
 la formaci\'on de estrellas masivas debe romper las nubes
 moleculares, mover el material, y comprimirlo en otras regiones para
 producir nuevas regiones de formaci\'on estelar.  As\'i, parece
 inevitable que este proceso c\'iclico resulte en un periodo de
 formaci\'on estelar extendido, el cual no representa la longevidad de
 una nube molecular individual.  Argumentamos que la formaci\'on
 estelar r\'apida indicada observacionalmente es tambi\'en m\'as
 f\'acil de entender te\'oricamente mque el escenario de contracci\'on
 cuasiest\'atica lenta v\'ia la difusi\'on ambipolar.}

\addkeyword{stars:formation}
\addkeyword{turbulence}

\begin{document}

\maketitle

\section{Introduction}

For many years the common picture of (particularly low-mass) star
formation was one in which protostellar cores quasi-statically
contract over timescales as long as 10~Myr until sufficient magnetic
flux has been removed by ambipolar diffusion to permit free-fall
collapse \citep{SAL87, Mouschovias91}.  This picture was motivated in
part by (a) the apparent need to avoid monolithic collapse of giant
molecular clouds, reducing thus the star formation efficiency (i.e.,
the fraction of a molecular gas mass that is converted into stars);
(b) the apparent need to reduce magnetic fluxes to the level necessary
to allow local gravitational collapse of magnetically supported clouds
\citep{Mestel_Spitzer56}; and (c) the old idea that all the forces (in
the ISM) should be in balance and the medium should have no net
acceleration \citep[see e.g.,][ Chap. 11]{Spitzer78}.  However, either
observational \citep{Jenkins_etal83, Bowyer_etal95, Jenkins_Tripp01,
Jenkins02, Redfield_Linsky04}, and numerical \citep{VS_etal03,
MacLow_etal05, Gazol_etal05} studies have found that the ISM is not in
pressure balance, but exhibits strong pressure fluctuations.
Equivalently, several studies on smaller scales have suggested that
star formation is not a quasistatic process, but generally rapid and
dynamic \citep[][ $=$HBB01]{Lee_Myers99, BP_etal99taurus, Elmegreen00,
Pringle_etal01, Briceno_etal01, HBB01}.  These studies seem to be in
better agreement with numerical simulations indicating that cloud
cores are more dynamic, have shorter lifetimes, and still a large
fraction of them may appear to have equilibrium density profiles
\citep{BP_etal03} and to be quiescent \citep{Klessen_etal05}.  In
response to these and other developments, several theoretical
approaches have been made to achieve shorter magnetic flux reduction
timescales, either by starting with a more nearly-critical field
strength, or enhancing ambipolar diffusion of magnetic fields through
turbulent motions, or both \citep{Ciolek_Basu01, Fatuzzo_Adams02,
Li_Nakamura04, Nakamura_Li05}.

In contrast to these investigations, \citet[][ $=$ TM04]{TM04} have
recently challenged the picture of rapid star formation, arguing that
the phase of cloud evolution prior to star formation has been ignored.
TM04 argue that by taking this potentially long phase of evolution
into account, the observations are consistent with the old scenario of
slow, magnetically-controlled star formation.  Additionally,
\citet[][$=$ MTK06]{MTK06} argue that galactic statistics are biased
and that the widths of spiral arms in external galaxies indicate
longer molecular cloud lifetimes.

In this paper we review the observational results that are in
disagreement with the view of  TM04  and  MTK06.  In
particular, we use the CO data from the survey by \citet{Dame_etal01},
in order to emphasize that, if clouds were living for many Myr before
they form stars, there should be much more clouds without young stars
and without internal structure than what is observed in the solar
neighborhood (within 1 kpc from the Sun).  We also identify problems
which contradict other observations often cited for long starless core
lifetimes and significant stellar age spreads.  We discuss
observations of protostellar core structure which by themselves
suggest that cores do not last for many free-fall times.  We stress
that in regions of massive star formation (where most stars form), the
destructive effects of massive stars on their environment cannot be
ignored when considering lifetimes of molecular clouds.  We also
review theoretical problems with the scenario of slow star formation,
emphasizing the importance of boundary conditions in understanding
star formation.  Finally, we suggest that the observations of external
galaxies are entirely consistent with {\em cycles} of cloud formation,
disruption, and reformation, rather than the lifetimes of single
molecular clouds.  We conclude that the observational evidence
supports the picture of rapid star formation, and that the rapid
formation scenario conforms well with recent numerical simulations.

\section{Statement of the Issue}\label{sec:theproblem}

The fundamental question we wish to address is whether star formation
is a dynamic, relatively rapid process,
or if it proceeds by slow, quasi-static evolution.  More precisely,
did the parcel of gas which ultimately ended up inside a star
experience long stretches of slow, quasi-static evolution in
near-equilibrium conditions, or was it always dynamic?  This question
is important because to the extent that magnetic fields dominate the
support of molecular clouds, as originally stated by TM04 and
references therein, one would expect slower evolution, addressable by
quasi-static, near-equilibrium calculations, but in which the origin
of cores is not addressed.  Conversely, if this evolution occurs on a
very few free-fall timescales, it suggests that magnetic field support
is not strong, that the formation of the cores themselves matter, and
that dynamic models of star formation are required.

The lifetime of a molecular cloud complex provides an {\it upper} bound to
the timescale over which an individual parcel of gas concentrates into
cores and then evolves to collapse, because core formation over areas
tens of pc or more distant need not be (and in general will not be)
coordinated perfectly in time.  Even with this caveat, the lifetimes
of molecular clouds in the solar neighborhood prove to be interesting.
As HBB01 pointed out, the average stellar population ages within these
clouds are about 1--2 Myr.
The free-fall timescale for an isothermal, uniform spherical
distribution of gas of density $\rho$ is \citep{Hunter62}

\begin{equation}
t_{\rm ff}  = \biggl( {3 \pi \over 32 G \rho }\biggr)^{1/2} \, ,
\end{equation}
which applied to molecular gas becomes

\begin{equation}
t_{\rm ff} \sim 3.4 \times n_{100}^{-1/2} \, {\rm Myr}\,,
\end{equation}
where $n_{100}$ is the density of molecular hydrogen in units of $100
\, {\rm cm}^{-3}$.  With a typical {\em average} volume density of a
giant molecular cloud (GMC) in the solar neighborhood of $\sim
50$~cm$^{-3}$ \citep[e.g.,][]{Blitz91}, it is clear that there must be
locally denser regions where the stars are forming to meet the
requirement of forming stars within 1-2 Myr; and many observations in
a variety of clouds clearly demonstrate that stars form in the much
denser regions present.  Furthermore, as only a small fraction of
substantial molecular clouds in the solar neighborhood do not harbor
young stars (HBB01) , and none of the local GMCs are devoid of star
formation \citep{Blitz91}, these denser regions must form quickly,
probably as part of the cloud formation process \citep[e.g.,][ for a
review]{Heitsch_etal05, VS_etal06, VS_etal06b, BP_etal06PPV}

It is important to recognize that the age spread in a stellar
association is not necessarily a useful constraint for the timescale
question posed here.  For example, the nearest B association to the
Sun, Sco-Cen, has an age spread of $\sim$~10--15~Myr, but the
molecular gas is confined to the adjacent Ophiuchus clouds, with a
stellar population having ages of $\sim$~1~Myr; the older regions are
devoid of molecular gas and thus have no continuing star formation
\citep[e.g.,][]{deGeus92}.  Moreover, stellar energy input through
photoioinization, winds, and supernovae can pile up gas in adjacent
regions and trigger later star formation \citep[e.g.,][]
{Elmegreen_Lada77}, creating an age spread of several Myr or more over
a volume of a few to tens of pc, as is seen in regions such as Sco-Cen
and Cep OB2 \citep{Patel_etal95}; but this age spread does {\em not}
represent the timescale of an individual parcel of molecular gas to
form stars.  Another example is that of Cep OB3, for which Burningham
\etal (2005) found some evidence for an age spread, but which lies
at the interface between an H II region and a dense molecular cloud,
suggesting multiple star forming epochs (e.g., Pozzo \etal 2003).  
The possible superposition of star-forming gas with previous
regions of star formation  should be kept in mind when considering observations
of distant extragalactic regions with limited spatial resolution (\S
\ref{sec:spiral}).

Molecular clouds in the solar neighborhood are especially useful in
addressing the timescale question posed above because the gas becomes
molecular at column densities such that self-gravity is important,
given pressures in the local interstellar medium \citep{Franco_Cox86,
Elmegreen82}.  They represent the regions of sufficient density to
form stars, and constitute only a fraction of the total gas nearby,
with a small volume filling factor.  In other regions where most of
the gas remains molecular, such as may be the case in the inner
molecular ring of the Galaxy, self-gravity need not be important
compared with external pressures in all regions containing molecular
gas, and the very definition of a molecular cloud is in question.

In summary, the issue we wish to address is how rapidly the matter of
stars is assembled.  Lifetimes of molecular clouds and ages of
associated stars provide upper limits to the local timescales of star
formation, the significance of these upper limits needs to be
considered carefully depending upon the circumstances.

\section{(Nearby) Molecular Cloud Lifetimes}\label{sec:mclifetimes}

\subsection{Ages of young stars in molecular clouds}

Despite the caveats about ages of stellar associations discussed
above, the ages of stars {\em within} molecular clouds provide the
essential observational result which implies rapid star formation.
This subject has been treated in detail by HBB01, Hartmann (2001), and
Hartmann (2003); here we briefly review some issues.

The fundamental starting point of any analysis is the recognition that
the median age of stars in nearby molecular clouds is $\sim 1-2$~Myr.
As Hartmann (2003) pointed out, this means that unless we are
observing at a special epoch in the history of local star formation --
an unattractive assumption -- the median age of star-forming molecular
clouds is also 1-2 Myr.

The more complicated issue is how to treat the apparent age spread.
As Palla \& Stahler (2000) showed, stars apparently older than about 3
Myr in most nearby star-forming regions are quite sparse; that is,
they form a small tail on the older end of the distribution (this is
implicit in the low median age; if the age spread for most of the
stars was much larger, the median age would have to be larger than 1-2
Myr).  The small number of stars in this tail of apparently older
stars implies a very low star formation rate compared with the
present.  As Hartmann (2003) pointed out, the Palla-Stahler model
implies that the molecular cloud was present for several Myr while
making very few stars.  Since most molecular clouds are actively
forming stars (see following section) at similar rates (Hartmann
2003), the Palla-Stahler model is inconsistent with observations.

A related issue has arisen recently concerning the massive Orion
Nebula Cluster (ONC).  Slesnick, Hillenbrand, \& Carpenter (2004)
carried out an infrared spectral survey of stars in the direction of
the ONC and found a population of M dwarfs with apparent ages $\sim
10$~Myr (assuming the same distance as the ONC).  In a related study,
Palla \etal (2005) argued that the Li depletion in two objects in the
ONC direction was consistent with an age $\sim 10$~Myr, as indicated
by their positions in the HR diagram; thus, as likely members of the
cluster, they demonstrate a significant age spread.
 
The older stars identified by Slesnick \etal (2004) and Palla \etal
(2005) raise the following question: what was the Orion Nebula region
like 10 Myr ago?  Was it really similar to its properties at the
present epoch, i.e. a dense concentration of $\gtrsim 5000$~$\msun$ of
molecular gas, while forming stars at an extremely low efficiency
compared with the current $\sim 30$\% (e.g., Hillenbrand \& Hartmann
1998)?  The authors are unaware of any comparable molecular region
which is not actively forming many stars; but according to the Palla
\etal (2005) picture, one would expect them to be common, as the ONC
would have to have most of its lifetime in the low-efficiency state.

It seems much more plausible to assume that the gas now in the ONC
region was much more dispersed, if it was indeed anywhere nearby at
that time.  A region of much lower densities and mass presumably would
produce stars at much lower rates, consistent with the observations.
One might imagine that small molecular clumps formed stars and then
dispersed in an early phase of accumulation of the Orion Nebula
region.

One must also emphasize that the membership of the older stars in the
ONC is far from certain.  The HR diagram displayed by Slesnick \etal
(2004) suggests a gap between the young and older stars, clearly
indicating two separate populations of stars, as would be the case if
the older stars were a foreground group.  Slesnick \etal discount the
idea of a foreground population, but Furesz \etal (in preparation) have found
some evidence for a population in the direction of the ONC with radial
velocities more consistent with those of the Orion 1a association than
with the ONC.  In the case of Palla \etal (2005), it is worth noting
that their theoretical depletion timescales are considerably shorter
than those of Baraffe \etal (1998).  As indicated, for example, in
Figure 2 of White \& Hillenbrand (2005), the Baraffe \etal
calculations would require much older ages than found by Palla \etal,
implying that they are a foreground population and thus appear younger
in HR diagrams assuming the same distance as the ONC.

In any event, it is clear that the most robust result is the median
age, at the peak of the apparent distribution of stellar ages.  As the
above discussion demonstrates, care must be used in interpreting the
tails of the age disitribution.

\subsection{Statistics of local molecular gas}

HBB01 pointed out that {\it most} nearby molecular clouds contain
young stars, that those stars are at most a few Myr old, while stellar
associations older than $5$~Myr contain no molecular gas.  Thus, star
formation {\em in the solar neighborhood} proceeds very quickly upon
molecular cloud formation and the timescales for star formation epochs
must not be much more than a few Myr, after which the star-forming gas
gets dispersed.

In contrast, TM04 argued that timescales of star formation estimated
from age distributions of the stars in molecular clouds ignore a
potentially lengthy period of pre-stellar cloud core evolution.
However, if there is a long lag time between molecular cloud formation
and star formation, as suggested by TM04, then on a statistical basis
{most} nearby molecular clouds should not be forming stars.  This is
observationally not the case, as already pointed out by HBB01.  To
expand upon our previous discussion, in Table~\ref{tabla:clouds} we
show the list of nearby (within 1 Kpc) molecular clouds, according to
\citet[][ see their Table~2] {Dame_etal87} Columns 1 and 2 indicate
respectively the name of the cloud, and its distance to the Sun, in
pc.  In column 3 we note whether the cloud is currently known to be
forming stars.  It should be noted that the total young star content
of the Cygnus and Aquila Rifts are under initial investigation
(R. Gutermuth, personal communication.) The ratio by number of
non-star-forming clouds to star-forming clouds is 7 to 14; by mass,
11.3 to 30.5 $\times 10^5 \msun$.  This is an upper limit to the
number or mass of non-star-forming clouds, because we know of no
careful search for young stars in clouds A, B, C, the Lindblad Ring,
and G317-4.  Given the typical ages of young stars in known
star-forming clouds of 1-2 Myr (HBB01, and references therein), the
extreme upper limit for the time between molecular cloud formation and
star formation is thus 1 Myr or less.  Thus, the argument of TM04 is
strongly at variance with the observations.

In an attempt to get around this problem, MTK06 tried a different
argument, using the steady-state equation
\be
{\tau_{\rm SF} \over \tau_{\rm MC} } = {N_{\rm NS} \over N_{\rm
tot}}\,, \label{eq:mtk}
\en
where $\tau_{\rm SF}$ is called the star-formation timescale,
$\tau_{\rm MC}$ is the molecular cloud lifetime, and $N_{\rm NS}$ and
$N_{\rm tot}$ are the number of molecular clouds without stars and the
total number of molecular clouds, respectively.  MTK06 adopt $N_{\rm
NS}/N_{\rm tot} \sim 0.1$ from the earlier tabulation of HBB01, and
then use $\tau_{\rm SF} =$~1~Myr from \citet{VS_etal05cores}, and then
conclude that $\tau_{\rm MC}$~$\sim$~10~Myr.

\begin{figure}[!t]
  \includegraphics[width=\columnwidth]{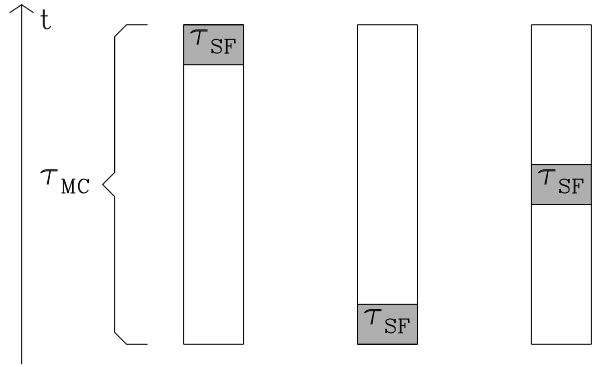} 
  \caption{Three illustrative cases in which the star formation
  episode $\tau_{\rm SF}$ is a small fraction (1/10th) of the total
  lifetime of the parental molecular cloud lifetime $\tau_{\rm MC}$,
  as suggested by TM04 and MTK06.  None of them is compatible with
  observations, as discussed in the text.}
 \label{fig:evolution}
\end{figure}

Although the numbers $N_{\rm NS}$ and $N_{\rm tot}$ in MTK06's
exercise are based on observational surveys (see HBB01), this result
is actually inconsistent with observations, and with the evolutionary
AD-based scheme depicted in Fig.~2a of TM04.  To explain this problem,
we have drawn in Fig.~\ref{fig:evolution} a similar figure to that of
TM04.  In this figure, we adopt the 10~Myr lifetime of MCs inferred by
MTK06.  In which moment of this 10~Myr lifetime occurs the star
formation? Judging from TM04's Fig.~2a, it should occur at the end of
the lifetime of the parental MC (see Fig.~\ref{fig:evolution}, left
panel).  But this is inconsistent with the cloud numbers used to infer
$\tau_{\rm MC}$~$\sim$~10~Myr above: precisely in this case there
should be 9 clouds {\it without} star formation per each cloud with
star formation.  In other words, in this case, the observed ratio
$N_{\rm NS}/N_{\rm tot}$ should be 0.9, not 0.1.

Another possibility to have a ratio $N_{\rm NS}/N_{\rm tot} \sim 0.1$
as the used by MTK06, is if the star formation event occurs at the
beginning of the molecular cloud lifetime, since only in this case,
there are few clouds without star formation.  This situation is drawn
in the middle panel of Fig.~\ref{fig:evolution}.  But this situation
has one observational and two theoretical problems.  In the first
case, we notice that there should be 10~Myr-old stars associated to
MCs, which is clearly in contradiction with observations, as it is
well known for 28 years now \citep{Herbig78}.  This is precisely the
so-called post-T Tauri problem that has given origin to the line of
thought that star formation should be fast.  As for the theoretical
problems, where is the (usually quoted as long) timescale needed for
ambipolar diffusion to allow inefficient star formation, if the stars
are formed right at the same time in which its parental MC is formed?
Moreover, why star formation stops after 1~Myr?  (if it goes on, then
$N_{\rm NS}/N_{\rm tot} \ne 0.1$, as used by MTK06).

One would be tempted at this point to argue that the situation should
fall in the middle of these two extremes (Fig.~\ref{fig:evolution},
right panel).  But this situation is also inconsistent with
observations.  According to our Table~\ref{tabla:clouds} and the
discussion above, either by mass, or by number, it is not true that
half of the nearby molecular clouds have formed stars while the other
half have not.  Moreover, as also we have mentioned above, the ages of
young stars associated to MCs are 1-2 Myr, while 5~Myr-old stellar
associations have no left molecular gas (HBB01).  Furthermore, as
above, what stopped the formation of stars after 1~Myr, in order to
keep the ratio $N_{\rm NS}/N_{\rm tot} \ne 0.1$?  As a conclusion, the
$\tau_{\rm MC}$~$\sim$~10~Myr estimation by MTK06 should to be wrong.

The problem with the MC ages estimated by MTK06 is that the timescale
$\tau_{\rm MC}$ is {\em not} the unknown variable; it is the {\em
known constraint} ($\lesssim$~3--5~Myr) from which the pre-star
formation lifetime of clouds is estimated.  We have no clear way to
determine lifetimes of non-star-forming molecular clouds
independently; they have no stars to provide age estimates.  Thus,
equation (\ref{eq:mtk}) should be used to determine $\tau_{\rm SF}$
from the other quantities, not the other way around.  If $N_{\rm NS}
\lesssim N_{\rm tot}/3$ (e.g., Table 1), then $\tau_{\rm MC}$ cannot
be much longer than the stellar ages $\sim 1 -2$~Myr in star-forming
molecular clouds.  This results in pre-star formation molecular cloud
timescales of order 1 Myr or less.

Additionally, the calculation MTK06 makes use of the value $\tau_{\rm
SF}\sim 1$ Myr obtained by \citet{VS_etal05cores} from numerical
simulations.  However, those simulations of start with an
already-formed uniform density molecular cloud, which is unlikely to
be a realistic initial condition.  Furthermore, by adopting periodic
boundary conditions, turbulent numerical simulations by different
groups (in particular the V\'azquez-Semadeni's quoted above) are
unable to address cloud dispersal - mass cannot be lost from the
system, and this is the essential process ultimately limiting
molecular cloud lifetimes.  It is inappropriate to use a simulation to
examine an issue which the simulation was never intended to address.

Another argument concerns the structure of molecular clouds.
According to TM04 (see their section 4, Fig.~3), molecular clouds
spend $\sim$~10--15~Myr before subcritical cores within them achieve
column density contrasts of 2--4, and only in the last few million
years dense cores become supercritical and collapse to form stars.
They argue that surveys miss a large part of the evolutionary phase of
molecular clouds because cores with column density contrasts of 2--4
are undetectable observationally.  If this model is applicable, there
should be a substantial fraction of local molecular clouds with column
density contrasts not larger than 2--4.
At issue is whether the lowest-density concentrations in molecular
clouds can be detected observationally, as this constitutes the longest
phase of ambipolar diffusion. 

To address this, we analyze the internal structure (named, the first
moments of the column density) of the local molecular clouds (see
Table 2).  We use $^{12}$CO data from \citet{Dame_etal01} because it
is a reasonably homogeneous and complete data set, while it provide us
a worst-case estimation of the actual column density: either by
depletion into grains, or by optically thick effects, any column
density inferred for the carbon monoxide is just a lower-limit value.
If even in this limit we found that most of the clouds exhibit column
density contrasts larger than 2--4, it will be clear that the argument
that cores evolving slowly are missed observationally is not
applicable.

Table \ref{tabla:clouds2} lists the properties of the local clouds.
Column 1 lists the name of the cloud, columns 2--4 list the maximum,
mean and rms CO column density, in K $\cdot$ km sec\alamenos 1.
Columns 5 and 6 list the ratio of maximum-to-mean and rms-to-mean CO
column densities\footnote{Note that the numbers quoted in Table
\ref{tabla:clouds2} have been computed by rejecting data below
3$\sigma$, with $\sigma=0.43$~K being the worst rms noise level of the
whole Galactic survey \citep{Dame_etal01}}.  Finally, column 7 shows
the width of the line integrated over the whole cloud, in km
sec\alamenos 1.  From this table we note that (a) that the typical
rms-to-mean column density is 0.75, indicating that the values of the
column density oscillate $\sim 1.5$ times around the mean column
density.  Thus, it is far from true that column density contrasts of
2--4 are missed in observational surveys; and (b) that the maximum
derived $^{12}$CO column density is {\it at least} 2.7 times larger
than the mean column density (for Chamaleon), and as large as 12 times
(for Orion A), with typical values of the order of $\sim$ 5.4 times
the mean column density.  This result show what is obvious by looking
at maps of any local molecular cloud: that every local molecular cloud
exhibits substantial substructure \citep[see, e.g., the maps published
by][ and a long etc.]{Tapia73, Dame_Thaddeus85, Dame_etal87,
Nyman_etal89, Mizuno_etal95, Dame_etal01, Onishi_etal02,
Straizys_etal03, Wilson_etal05}.  Thus, even in this worst-case
scenario, there are no MCs without large-column density contrasts, as
will be the case if MCs spend a large fraction of their lives evolving
from subcriticallity to supercriticallity.  We emphasize that while
there are uncertainties in translating column densities into volume
densities, $^{12}$CO is usually so optically thick that it strongly
underestimates the true variations in column density.

In any event, the timescale of generating substructure within any
molecular cloud cannot be longer than the total lifetime of the cloud
itself; and as pointed out earlier, this cannot be a long timescale in
the solar neighborhood.

\subsection{H$_2$ formation timescales}

An important question for any model of star formation is how long does
it takes to a parcel of interstellar atomic gas to become molecular.
Traditionally, this timescale has been thought to be large, since the
formation of molecular hydrogen in the ISM occurs in timescales given
by $t_{\rm H_2 form} \simeq 10^9 yr/n$, where $n$ is the number
density in cm\alamenos 3 \citep[e.g.,][]{Jura75}, meaning that a GMC
with mean density of the order of $n\sim 100$~cm\alamenos 3 has spent
$\sim$~10~Myr in the transition from atomic to molecular.  This
timescale is favored by chemical evolution models at constant density
by, e.g., \citet[][=GL05]{Goldsmith_Li05}.  These authors have argued
that lifetimes of molecular cloud cores are long, based on detections
of small amounts of cold H I in dense cores, and interpreting the
H~I/H$_2$ ratios in terms of chemical evolution at constant density in
regions shielded from the interstellar radiation field.  GL05 found
``minimum'' lifetimes of 3-20 Myr, much longer than estimated from
stellar ages.  However, there are substantial uncertainties in the
chemical rates and the physical model employed which render these
estimates suspect.

As GL05 themselves state in \S~6.1 of their paper, ``Combining the
uncertainties in the various factors, $k^\prime_{{\rm H}_2}$ (the rate
for forming H$_2$) may differ from its nominal value by a factor of 5.
This directly affects the H$_2$ formation timescale and the steady
state H~I density, both of which vary as $1/k^\prime_{{\rm
H}_2}$.''  As an example of this, Bergin et al. (2004) used a
sticking probability of 1 instead of 0.3 adopted by GL05, making the
timescales shorter by a factor of three.  As a further illustration of
uncertainties, GL05 note that if they had used C$^{18}$O densities
instead of C$^{13}$O densities, the timescales would be reduced by a
factor of two.

Furthermore, the physical model used by GL05 is probably unrealistic.
The shock model of Bergin et al. (2004) for molecular cloud formation
showed that H$_2$ formation can be nearly complete before the cloud
becomes ``molecular" in the sense of having significant CO.  This
means that much of the evolutionary time in the GL model should be
attributed to the atomic phase, not the molecular cloud phase.
Moreover, as \citet{Glover_MacLow06} have recently shown, supersonic
turbulence can enhance the production of molecular hydrogen. In fact,
these authors show that the low density regions ($n < 300$~cm\alamenos
3) of numerical simulations of molecular clouds exhibit more H$_2$
than the expected amount of H$_2$ if the gas were in
photo-dissociation equilibrium.  The physical mechanism is simple: a
large fraction of the H$_2$ found at low densities was actually
rapidly formed at higher densities, in gas with $n > 1000$ cm\alamenos
3, but subsequently transported to low densities by the advection of
the turbulent velocity field \citep{Glover_MacLow06}.

\subsection{Dispersal}\label{dispersal:sec}

An important constraint on molecular cloud lifetimes clearly comes
into play when massive stars are present.  Such stars are very
destructive to their natal clouds \citep[see][for a review]
{BP_etal06PPV}.  For example,
\citet{Leisawitz_etal89} found that clusters older than ~10 Myr do not
have associated with them molecular clouds more massive than a few
times \diezala 3 $\msun$.  More recently,
in the nearest B association, Scorpius-Centaurus, which consists of
stars with ages of $\sim$~5--15~Myr, molecular gas is not present;
instead, one can observe large H I shells around the three
sub-concentrations --Lower Centaurus-Crux, Upper Centaurus-Lupus, and
Upper Scorpius-- which are probably the result of the dispersal of
association gas by stellar winds and supernovae \citep{deGeus92}.  As
\citet{deGeus92} showed, the action of a single supernova would be
sufficient to remove the gas in the 5~Myr-old Upper Scorpius
sub-association \citep{Preibisch_Zinnecker99}, while the molecular gas
at the eastern end of the region remains as the Ophiuchus cloud, with
young stars (of ages $\sim$~1~Myr or less) and forming protostars.

As another example, the Cep OB2 association consists of a central
10~Myr-old cluster, NGC 7160, surrounded by a partial ring/bubble of
radius $\sim 50$~pc of molecular and atomic gas \citep{Patel_etal95},
and with recent star formation within the molecular bubble.  As
\citet{Patel_etal95} showed, this extended distribution of gas is
consistent with being blown out by stellar winds and supernovae from
NGC 7160.  On a smaller scale, the $\sim 4$~Myr-old cluster Trumpler
37, lying near the rim of the bubble, has itself a blown-out region of
several pc in radius; the central O7 star has driven out material due
to the pressure of the H II region (IC 1396), which has called a halt
to star formation within the bubble except in small globules of
molecular gas which contain $\sim 1$~Myr-old stars
\citep{Sicilia-Aguilar_etal04, Sicilia-Aguilar_etal05}.

We also note two other examples of rapid dispersal. The O9.5V star
$\sigma$ Ori   
has dispersed most of the gas surrounding its low-mass stellar cluster
by an age of $\sim 2.5$~Myr \citep{Sherry et al. 2004}, and in
the $\lambda$ Ori star-forming region, gas has been cleared out
and star formation has ceased out to a distance of 20 pc over a timescale
of 5-6 Myr \citep{Dolan_Mathieu01}.

In summary, it is very likely that giant molecular clouds in the solar
neighborhood are disrupted by the energy input of their massive stars
over timescales of order 5~Myr or perhaps even less in some cases,
explaining why regions of ages $5-10$~Myr, such as Scorpius-Centaurus
and Orion 1a, are devoid of molecular gas and ongoing star
formation. This addresses another ``problem'' that magnetically-slowed
star formation was supposed to solve, specifically, the inefficiency
of molecular gas in forming stars.  Star formation can be relatively
rapid and still result in a low efficiency as long as molecular clouds
are dispersed rapidly.  It is less clear what happens to low-mass
molecular clouds; they may be blown away by their own outflows, or
nearby supernovae may also play a role.

\section{Starless Cores}\label{sec:starless}

\subsection{Statistics}

Another question is how long protostellar cores last before collapsing
to form stars.  The lifetimes for starless cores have been estimated
by several authors using, again, steady state:
\begin{equation}
t_{\rm SC} = {N_{\rm SC} \over N_p} t_p\,, \label{eq:tcore}
\end{equation}
where $t_{\rm SC}$, is the lifetime of the starless cores, $N_{\rm
SC}$ is the number of starless cores, $N_p$ is the number of
protostars or embedded (heavily-extincted) young stars, and $t_p$ is
the corresponding lifetime.  TM04 argue that $t_p$ is only estimated
theoretically, leading to uncertainty in equation
(\ref{eq:tcore}). However, one can estimate the protostellar lifetime,
again assuming an approximate steady state, by using the ratio of
protostars or embedded sources to T Tauri stars, and using estimated T
Tauri ages.  In Taurus, the ratio of protostars (Class I objects) to T
Tauri stars is approximately 1:10 \citep{Kenyon_etal90,
Kenyon_etal94}; given an average age in of Taurus stars of roughly 2
Myr \citep{Kenyon_Hartmann95, Hartmann03}, this means that $t_p
\sim 2 \times 10^5$~yr.  This result is consistent with theoretical
expectations for free-fall collapse \citep[e.g.,][]{SAL87}.  Thus,
$t_p$ is so short that $t_{\rm SC}$ cannot be many millions of years
unless $N_{\rm SC}$ is more than an order of magnitude larger than
$N_p$.

As TM04 note, the studies of \citet{Lee_Myers99} and
\citet{Jijina_etal99} derive estimates of the lifetime of the starless
core phase between 0.1 and 0.5~Myr, inconsistent with slow contraction
controlled by (substantial) ambipolar diffusion.  TM04 in contrast
cite results from \citet{WardThompson_etal94} and
\citet[][$=$ JWT00]{Jessop_WardThompson00} which suggest lifetimes of
order $10^7$ yr for lower-density cores.  However, the
\citet{Lee_Myers99} and \citet{Jijina_etal99} core surveys are much
more heavily weighted toward nearby regions which have been the
subject of much more extensive ground-based optical and infrared
studies than the JTW00 study focused on generally more distant and far
less well-studied regions, and used only the IRAS point source catalog
to find embedded members rather than often more sensitive near- to
mid-infrared ground-based studies.  IRAS source counts of the JTW00
regions, at least half of which lie at distances of 300--800 pc, are
very likely to be significantly incomplete.

To illustrate the problem, the median 60 $\mu$m IRAS fluxes in Taurus
are $\sim 6$~Jy for protostars (Class I) and $\sim 1.4$~Jy for
accreting (Class II) T Tauri stars.  Although a lower limit of 0.4 Jy
is claimed by JWT00 at $60 \mu$m, they detect no source fainter than
about 1 Jy at this wavelength.  At this limit, about half of the
Taurus protostars would have been missed at distances $d \gtrsim
6^{1/2} \times 140$~pc $\sim 350$~pc; almost all of the T Tauri stars
would be undetectable.  This estimate assumes that the flux decrease
due to distance is the only difficulty, but in fact there are
additional problems due to the large beam sizes of IRAS, which lead to
increasing source confusion and problems with background subtraction
with increasing distance.  The result is that IRAS source counts in
any but the closest star-forming regions are likely to underestimate
stellar source populations by large factors, suggesting that the
\citet{Lee_Myers99} and \citet{Jijina_etal99} results are more
reliable for stellar statistics.
  
As a specific example, \citet{Reach_etal04} used the Spitzer Space
Telescope to detect 8 embedded sources (Class 0/I) in a small field
centered on a single globule in Tr 37; at this distance \citep[$\sim
900$~pc,][]{Contreras_etal02}; only one of these sources was detected
with IRAS.  \citet{Young_etal04} demonstrated that L1014, a dense core
previously thought to be starless, actually shows evidence of an
embedded source in higher-sensitivity Spitzer observations, leading
them to suggest that traditional estimates of pre-stellar core
lifetimes may be overestimated.

Equation (\ref{eq:tcore}) assumes that every starless core will end up
forming a star(s).  However, cores may be disrupted by shocks as well
as formed.  Numerical simulations of turbulent molecular clouds show a
population of cores that do not end up collapsing, but re-dispersing
on a fraction of the ambipolar diffusion timescale \citep[e.g.,
][]{VS_etal05cores, Nakamura_Li05}.  Thus, even if these simulations
included ambipolar diffusion, there may not be enough time for AD to
operate and allow the core to become supercritical.  In this
connection it is important to note that the JWT00 cores on the average
have considerably less column density and are more extended than the
\citet{Lee_Myers99} and \citet{Jijina_etal99} cores, and thus may be
less likely to eventually collapse than the objects in the latter two
surveys.

TM04 suggest that because of the difficulty of recognizing
low-density, slowly-evolving cores, some studies may have undercounted
$N_{\rm SC}$ and thus greatly underestimated the lifetimes of starless
cores.  However, molecular cloud cores must be situated in molecular
clouds by definition; and if clouds do not spend a long time before
forming stars, cores cannot spend a long time evolving before forming
stars.

Given the potential for source count incompleteness in some surveys
and the possibility that not all cores will collapse to form stars, it
appears that equation (\ref{eq:tcore}) provides an upper limit for the
lifetime estimate.  The much deeper samples that will be obtained
using the Spitzer Space Telescope in the near future should strongly
reduce source incompleteness \citep[e.g., ][]{Young_etal04}.

\subsection{Core Morphology}

If protostellar cores are really quasi-equilibrium objects, one might
expect them to be more regular in shape.  Detailed images of cores
show that many have extended and not entirely regular or smooth shapes
\citep[e.g.,][]{Myers_etal91, Bacmann_etal00, Steinacker_etal05},
which as \citet{Myers_etal91} pointed out, suggests problems for
equilibrium models.  Irregular cores suggest that they are {\it not}
objects which have lasted for several free-fall times, but instead
have transient structures, as seen in numerical simulations \citep[][
see Ballesteros-Paredes et al. 2006 for a review]{BP_ML02,
Gammie_etal03, Li_etal04}.

Another long-standing problem for theories in which magnetic fields
strongly constrain core structure is the general finding that cores,
irregular as they are, are more nearly prolate than the naturally
oblate structure expected for compression along magnetic field lines.
This conclusion has been reached by different analyses assuming random
distributions on the sky \citep{Myers_etal91, Ryden96}.  In the case
of Taurus, the strong tendency of the cores to be elongated not
randomly but along filaments makes the argument even stronger for more
prolate than oblate objects \citep{Hartmann02}.
\citet{Curry_Stahler01} conducted a careful study of the structure and
equilibria of prolate cores embedded in filaments.  Although
equilibrium solutions can be found, they noted that the structure of
the field lines in such solutions were dynamically unstable in
laboratory plasmas, and suggested that their prolate states would
rapidly transform into lower-energy configurations.

Perhaps an even greater problem is the global nature of gravity in
molecular clouds in general and dense filaments in particular.  In
their solutions, \citet{Curry_Stahler01} either assumed no external
gravitating mass or a vanishing gravitational force at specified
distances along the filament, as for example with an infinite chain of
equal cores.  However, as \citet{Burkert_Hartmann04} have argued, it
is extremely difficult to set up an equilibrium condition for a finite
filament of many Jeans masses; even with supporting turbulent motions
or rotation or expansion, it is difficult to avoid gravitational
collapse somewhere.  Equilibrium models of cores ignore the important
and unavoidable questions of just how one prevents
gravitationally-driven motions arising from large scale mass
distributions, motions which can eliminate local equilibria
\citep{Burkert_Hartmann04}.

\subsection{Mass to Flux Ratios}\label{m2f:sec}

\citet{Ciolek_Basu01},  TM04, and  MTK06  point out that
the ambipolar diffusion timescale, and thus the starless core
evolutionary timescale in their picture, is not a constant but depends
upon the initial mass-to-magnetic flux ratio.  If in a region of star
formation, one adopts an initial mass-to-flux ratio close to critical,
ambipolar diffusion in small subregions can be fast enough that it
does not significantly lengthen the timescale of core collapse.
Moreover, if the {\em average} mass-to-flux ratio in a molecular cloud
is near critical, assuming variations in this ratio under realistic
conditions will yield some supercritical regions as well as
subcritical areas; it would seem likely that the supercritical regions
would be easier to condense and then faster to collapse than the
subcritical regions.

It is also worth emphasizing that, under the assumption of
subcriticallity, gravity {\em cannot} be responsible for assembling the
mass of the core, because by definition the magnetic forces are
stronger than the opposing gravitational forces; non-gravitational
flows must then converge to make the mass concentrations.  Given the
rapidity with which star formation follows molecular cloud formation
in the solar neighborhood, it is not clear that one can consider
equilibrium or quasi-equilibrium models for these structures without
showing that the flows responsible for forming the cores in the first
place do not distort or buffet the core; and especially, that the
flows are not gravitationally-driven, which would imply
supercriticallity.

Recent numerical simulations of clumps in subcritical boxes have been
performed by \citet{VS_etal05cores} and \citet{Nakamura_Li05}.
\citet{VS_etal05cores} found that in subcritical boxes, only
a minority of moderately-gravitationally bound clumps form, but they
are re-dispersed by the large-scale supersonic turbulence on
timescales smaller than the local ambipolar diffusion timescale,
suggesting that only a small fraction of cores can be {\it marginally}
affected by ambipolar diffusion to increase their mass-to-flux ratio
and eventually collapse.

\citet{Nakamura_Li05} found that subcritical cores formed in a
turbulent medium can have shortened ambipolar diffusion timescales and
thus collapse rapidly \citep[see also][]{Fatuzzo_Adams02}.  However,
their simulations suggests only a {\it limited} importance for
ambipolar diffusion in the collapse of subcritical cores for the
following reasons: (a) their simulations, although initially at Mach
10, are decaying, and thus, the flow spent most of the time at low
Mach numbers, $\cal M \sim$ 2--3.  By simple inspection of column 7 in
Table 2, the velocity dispersion reported for nearby molecular clouds
is at least 10 times larger than the sound speed ($\sim$ 0.2 km
sec\alamenos 1 if clouds are nearly isothermal at $\sim$~10~K).  In
other words, the simulations by \citet{Nakamura_Li05} assume a much
less violent medium than actual molecular clouds. (b) In a turbulent
medium, the timescale for core formation is much shorter than
AD-mediated contraction because a core forms in the turbulent crossing
time for the larger scale from which it gathers its mass, rather than
on the AD-timescale \citep{Li_Nakamura04, Nakamura_Li05,
VS_etal05cores, BP_etal06PPV}. (c) \citet[][ see also Li \& Nakamura
2004]{Nakamura_Li05} found their best results for marginally
subcritical clouds, with mass to flux ratios only 20\% smaller than
the critical value.  Making clouds only slightly subcritical rather
than strongly subcritical reduces the ambipolar diffusion timescale by
a similar ratio, even without turbulent enhancement.  Again, if clouds
are close to overall criticality, one wonders whether fluctuations in
the flux to mass ratio would result in initially supercritical density
enhancements which could collapse without flux loss.

Furthermore, it is far from clear that subcritical boxes are a good
representation of global molecular cloud conditions.
\citet{Bertoldi_McKee92} and \citet{McKee_etal93} suggested that, as a
whole, molecular cloud complexes are magnetically supercritical, and
\citet{Bertoldi_McKee92} extended this statement of supercriticality
to most clumps.  \citet{Nakano98} argued that cores must also be
generally supercritical if they are objects of higher-than-average
surface density (which of course is generally expected of cores).

If molecular clouds are, in fact, supercritical, then even if
subcritical regions exist within such clouds, there must be even more
supercritical regions within the cloud by definition; and one would
expect star formation to occur fastest, and thus preferentially, in
such especially supercritical regions.  Numerical simulations in which
the boundary conditions do not demand subcriticality support Nakano's
argument; subcritical regions are low-density, while the
highest-density regions tend to be supercritical (see HBB01).
Slowly-evolving, quasi-static, significantly subcritical cloud cores
do not seem to appear in numerical simulations with supersonic
turbulence characteristic of molecular clouds \citep[Mach numbers of
the order of 10, see, e.g.,][]{MacLow_etal98, Stone_etal98,
BP_etal99scalo, Heitsch_etal01, Li_Nakamura04, VS_etal05cores}.

Finally, the traditional picture of slow, quasi-static contraction of
initially subcritical molecular cloud cores does not address how such
cores are formed in the first place.  The numerical simulations,
whatever their limitations, illustrate an important point: it does not
seem to be straightforward to make and maintain a subcritical,
quasi-static cloud core, confined by external pressure, if that
external pressure is both anisotropic and time-dependent.  It is
important to consider the problem of core formation and their
maintenance within a supersonic, turbulent medium.

\section{Spiral Arms and Star Formation in other Galaxies}\label{sec:spiral} 

TMK06 argue that observations on larger scales than the solar
neighborhood need to be considered to understand timescales of star
formation.  For one thing, they argue that the local census of
molecular clouds and star formation is highly biased because most
regions are embedded just downstream of the spiral arm shock, due to
the difference in rotation and spiral pattern speeds, and that these
regions are difficult to observe.

In HBB01 we limited our direct conclusions to the solar neighborhood
within about 1 Kpc for observational completeness reasons (see Table 1
and \S\ref{sec:theproblem}).  This constitutes an interarm region in
the galaxy \citep[e.g., ][]{Taylor_Cordes93}.  However, this does not
detract from the fact that the long-lived molecular cloud scenario
does not hold locally.  Furthermore, simply because more distant,
confused regions have not been studied adequately does not
automatically mean that they refute the concept of dynamic star
formation.

Another argument made by TMK06 is that the external galaxies M51 and
M81 show a spatial separation between the dust lanes in spiral arms --
which correspond with the peak of the CO emission -- and the peak in
H$\alpha$ emission \citep[citing, e.g., ][]{Vogel_etal88}.  Using
estimated differences between rotation and pattern speeds, they
interpret this as a time lag between the formation of molecular clouds
and the formation of stars on the order of 10~Myr.

However, this interpretation demands that the molecular clouds
maintain their identity over size scales of 100 pc or more and
timescales of $\gtrsim 10$~Myr.  Specifically, this interpretation
{\em assumes} that the molecular gas does not get dispersed by the
action of stellar photoionization, winds, and supernovae, and then
gets concentrated in other locations to make new clouds which make
additional stars.  Suppose that a first generation of stars forms
within the CO clouds just after passing through the spiral shock.  The
most massive of these stars will start disrupting their environments
as they form H II regions (note: the very process of forming the H II
region will shut off {\em local} star formation; see
\S\ref{sec:theproblem}).  The expansion of the H II regions, and
eventually supernovae, will compress gas to make secondary generations
of stars with additional H~II regions, etc. --just as observed in
nearby regions such as Cep OB2 (\S\ref{sec:theproblem}).  Thus the
spiral shock wave represents the beginning of the (locally rapid) star
formation process by concentrating gas.  As time goes on, gas is
disrupted (H II regions) and flows make new concentrations which make
new generations of stars with associated H II regions.  In this
dynamic picture, the peak of the H II region emission extends further
downstream than the highest concentration of molecular gas because of
succeeding phases of expansion and compression, with succeeding local
events of star formation, which eventually slow down as gas gets
dispersed by stellar energy input and eventually also due to changes
in the gravitational potential.  The lifetimes of H II regions, if
more than 1--2~Myr, can also add to the distance downstream of peak H
II emission; the longer H II regions last (this is the timescale of
dispersing the local ionized gas), the further downstream they will
appear.

It is important to emphasize, as found by \citet{Vogel_etal88}, that
in the relevant regions of M51 {\em most} of the gas is molecular, not
atomic; these authors conclude that {\em most} of the gas is in the
interarm region, not in the ``arms''.  This raises the question of
just where a molecular ``cloud'' begins and ends
(\S\ref{sec:theproblem}).  And it seems inevitable that the giant H II
regions must have molecular gas around them --since they formed
relatively recently-- gas which does not show up in the interferometer
maps.  It seems quite possible that the CO arms in
\citet{Vogel_etal88}, represent the first concentration of gas
downstream from the spiral shock, but not the densest clumps which are
the true progenitors of the H II regions.  Thus, it is far from clear
that the observations of spatial displacement between CO spiral arms
and H II regions demands a long lifetime of molecular clouds as a
physical entity, rather than as a complex region with locally rapid
evolution.  And in any event, in terms of the issue we have posed here
-- the evolution timescale of parcels of self-gravitating gas
(\S\ref{sec:theproblem}) -- the very formation of an H II region
implies that {\em local} star formation timescales are relatively
short, independently of whatever is deemed to be a molecular cloud or
cloud complex.

\section{Conclusions}

The scarcity of (well-studied) molecular clouds without star formation
indicates that the time lag between cloud formation and star formation
in the solar neighborhood is short.  Statistical estimates of
pre-stellar core lifetimes in well-characterized star forming regions
indicate that the pre-stellar phase is short, which also supports
rapid star formation.  Age spreads in well-studied and
carefully-analyzed star-forming regions are at most a few Myr.
Theoretical studies of cloud pressure balance, core formation, and
core evolution in turbulent gaseous clouds are consistent with the
observational evidence for rapid star formation, possibly because
magnetically supercritical or at least critical conditions are
generally applicable.

\acknowledgements

We thank Tom Dame for providing us the CO data from his Milky Way
2001's survey.  This work was supported in part by DGAPA-UNAM grant
IN110606 to J. B.-P.\ and NASA grant NAG5-9670 and NAG5-13210 to L. H.
This work has made extensive use of NASA's Astrophysics Abstract Data
Service and LANL's astro-ph archives.

\begin{table*}[!t]\centering
  \newcommand{\DS}{\hspace{6\tabcolsep}} 
  \setlength{\tabnotewidth}{0.9\textwidth}
  \setlength{\tabcolsep}{1.33\tabcolsep}
  \tablecols{5}
  \caption{Molecular Clouds within 1~kpc from the Sun}\label{tabla:clouds} 
  \begin{tabular}{c c c c c}  
   \toprule
	  {Cloud}
	& {Distance from Sun}
	& {Star Formation?}
	& {Mass}
	& {Ref}
\\
	 
	& {[pc]}
	&
	& {[$10^{5} M_\odot$]}
	&\\

\midrule
	 {Aquila Rift}
	&{270}
	&{yes}
	&{2.7}
	&{1,2,4}
\\
	 {Cloud A}
	&{500}
	&{?}
	&{0.4}
	&{1}
\\
	 {Cloud B}
	&{300}
	&{?}
	&{0.4}
	&{1}
\\
	 {Cloud C}
	&{500}
	&{?}
	&{0.3}
	&{1}
\\
	 {Vul Rift}
	&{400}
	&{linked to VulOB1}
	&{0.8}
	&{1,3}
\\
	 {Cyg Rift}
	&{700}
	&{yes}
	&{8.6}
	&{1,4}
\\
	 {Cyg OB7}
	&{800}
	&{yes}
	&{7.5}
	&{1}
\\
	 {Lindblad Ring}
	&{300}
	&{?}
	&{1.6}
	&{1}
\\
	 {-12 km/sec}
	&{800}
	&{?$^1$}
	&{8.7}
	&{1}
\\
	 {Cepheus}
	&{450}
	&{yes}
	&{1.9}
	&{1}	
\\
	{Taurus}
	&{140}
	&{yes}
	&{0.3}
	&{1}
\\
	 {Perseus}
	&{350}
	&{yes}
	&{1.3}
	&{1}
\\
	 {Monoceros}
	&{800}
	&{yes}
	&{2.8}
	&{1}
\\
	 {Orion}
	&{500}
	&{yes}
	&{3.1}
	&{1}
\\
	 {Vela}
	&{425}
	&{yes}
	&{0.8}
	&{1}
\\
	 {Chamaleon}
	&{215}
	&{yes}
	&{0.1}
	&{1}
\\
	 {Coalsack}
	&{175}
	&{no}
	&{0.04}
	&{1}
\\
	 {G317-4}
	&{170}
	&{?}
	&{0.03}
	&{1}
\\
	 {Lupus}
	&{170}
	&{yes}
	&{0.3}
	&{1}
\\
	 {$\rho$ Oph}
	&{165}
	&{yes}
	&{0.3}
	&{1}
\\
	 {R CrA}
	&{150}
	&{yes}
	&{0.03}
	&{1}
\\
\bottomrule
\tabnotetext{1} {Dame et al. 1987}	 
\tabnotetext{2} {Straizys et al. (2003), A\&A, 405, 585. }
\tabnotetext{3} {Fresneau \& Monier (1999) AJ, 118, 421}
\tabnotetext{4} {Gutermuth 2006, personal communication}
\tabnotetext{a}{The ``$-12$ km/sec'' cloud is a fuzzy cloud in the
second quadrant which may very well be associated to the Gould Belt,
which is well known to have stars.}

\end{tabular}
\end{table*}


\begin{table*}[!t]\centering
  \newcommand{\DS}{\hspace{6\tabcolsep}} 
  \setlength{\tabnotewidth}{0.9\textwidth}
  \setlength{\tabcolsep}{1.33\tabcolsep}
  \tablecols{7}
  \caption{CO column density contrast and velocity dispersion of local
MCs}\label{tabla:clouds2} 
\begin{tabular}{c c c c c c c }
\toprule
{Cloud} &{$N_{\rm max}$} &{$N_{\rm mean}$} &{$N_{\rm rms}$} &{$N_{\rm
 max}/N_{\rm mean}$} &{$N_{\rm rms}/N_{\rm mean}$} &{$\Delta v$}  \\
         &{[K km sec\alamenos 1]} &{[K km sec\alamenos 1]} &{[K km
       sec\alamenos 1]} &{} &{} &{[km sec\alamenos 1]}  \\
\midrule
	 {Aquila Rift} &{57.19} &{9.58} &{6.34} &{5.97} &{0.66} &{}
\\
	 {Cloud A} &{28.23} &{6.15} &{4.86} &{4.59} &{0.79} &{5}
\\
	 {Cloud B} &{27.81} &{5.74} &{4.06} &{4.85} &{0.71} &{3.5}
\\
	 {Cloud C} &{34.94} &{5.88} &{5.45} &{5.94} &{0.93} &{4}
\\
	 {Vul Rift} &{23.25} &{5.41} &{3.35} &{4.30} &{0.62} &{5}
\\
	 {Cyg Rift}
	&{81.33}
	&{12.79}
	&{12.19}
        &{6.36}
	&{0.95}
	&{13}
\\
	 {Cyg OB7}
	&{48.45}
	&{7.95}
	&{6.17}
        &{6.09}
	&{0.78}
	&{7}
\\
	 {Lindblad Ring}
	&{33.24}
	&{5.08}
	&{3.71}
        &{6.54}
	&{0.73}
	&{$>$ 7 (r)}
\\
	 {$-12$ km/sec}
	&{56.98}
	&{6.12}
	&{4.98}
        &{9.30}
	&{0.81}
	&{$>$ 4 (l)}
\\
	 {Cepheus}
	&{27.49}
	&{5.92}
	&{4.48}
        &{4.65}
	&{0.76}	
	&{4}
\\
	{Taurus}
	&{27.51}
	&{7.57}
	&{4.59}
        &{3.63}
	&{0.61}
	&{3}
\\
	 {Perseus  OB2 1}
	&{65.23}
	&{10.81}
	&{10.06}
        &{6.04}
	&{0.93}
	&{$>$ 3.5 (r)}
\\
	 {Perseus OB2  2}
	&{38.6}
	&{7.22}
	&{4.40}
        &{5.29}
	&{0.61}
	&{3}
\\
	 {Monoceros OB1}
	&{73.75}
	&{11.82}
	&{10.24}
        &{6.24}
	&{0.87}
	&{6}
\\
	 {Orion A}
	&{159.71}
	&{12.75}
	&{14.16}
        &{12.53}
	&{1.11}
	&{6--7}
\\
	 {Orion B}
	&{123.23}
	&{13.21}
	&{15.51}
        &{9.33}
	&{1.17}
	&{5}
\\
	 {Monoceros R2}
	&{57.85}
	&{8.73}
	&{6.68}
        &{6.63}
	&{0.77}
	&{4}
\\
	 {Vela Sheet}
	&{17.14}
	&{4.40}
	&{2.80}
        &{3.90}
	&{0.64}
	&{ $>$ 5 (l)}
\\
	 {Chamaleon}
	&{20.91}
	&{7.54}
	&{4.05}
        &{2.77}
	&{0.54}
	&{5}
\\
	 {Coalsack}
	&{14.35}
	&{4.65}
	&{2.69}
        &{3.09}
	&{0.58}
	&{5}
\\
	 {G317-4}
	&{16.41}
	&{5.29}
	&{3.06}
        &{3.19}
	&{0.58}
	&{3}
\\
	 {Lupus}
	&{23.28}
	&{6.54}
	&{4.43}
        &{3.56}
	&{0.68}
	&{4}
\\
	 {$\rho$ Oph 1}
	&{51.34}
	&{8.34}
	&{6.83}
        &{6.16}
	&{0.82}
	&{4}
\\
	 {$\rho$ Oph 2}
	&{22.13}
	&{6.96}
	&{4.53}
        &{3.18}
	&{0.65}
	&{3.5}
\\
	 {R CrA}
	&{28.71}
	&{6.20}
	&{5.19}
        &{4.63}
	&{0.84}
	&{3}
\\
\bottomrule
\tabnotetext{1} {Dame et al. 1987}	 
\tabnotetext{2} {Straizys et al. (2003), A\&A, 405, 585. }
\tabnotetext{3} {Fresneau \& Monier (1999) AJ, 118, 421}

\end{tabular}
\end{table*}

\end{document}